\documentstyle[12pt]{article}

\input{epsf}
\def\Journal#1#2#3#4{{#1} {\bf #2}, #3 (#4)}

   \def\unlock{\catcode`@=11}

   \unlock

   \def\gsim{\mathrel{\mathpalette\@versim>}}
 
   \def\lsim{\mathrel{\mathpalette\@versim<}}

   \def\@versim#1#2{\vcenter{\offinterlineskip
        \ialign{$\m@th#1\hfil##\hfil$\crcr#2\crcr\sim\crcr } }}

\def\NPB{{\em Nucl. Phys.} B}
\def\PLB{{\em Phys. Lett.}  B}
\def\PRL{\em Phys. Rev. Lett.}
\def\PRD{{\em Phys. Rev.} D}

\def\ra{\rightarrow}

\def\al{\alpha}

\def\be{\begin{equation}}
\def\ee{\end{equation}}
\def\bea{\begin{eqnarray}}
\def\eea{\end{eqnarray}}

\begin{document}

\topskip 2cm 
\begin{titlepage}

\hspace*{\fill}\parbox[t]{4cm}{EDINBURGH 96/11\\ July 1996}

\vspace{2cm}

\begin{center}
{\large\bf Theory of Double Hard Diffraction} \\
\vspace{1.5cm}
{\large Vittorio Del Duca} \\
\vspace{.5cm}
{\sl Particle Physics Theory Group,\,
Dept. of Physics and Astronomy\\ University of Edinburgh,\,
Edinburgh EH9 3JZ, Scotland, UK}\\
\vspace{1.5cm}
\vfil
\begin{abstract}
\noindent
In this review talk I consider the physics of 
rapidity gaps between two jets at hadron colliders, as a preliminary 
investigation toward understanding the production of a Higgs boson 
via weak boson fusion at the LHC.
\end{abstract}

\vspace{2cm}

{\sl To appear in the Proceedings of\\ the XI Topical Workshop
 on P-Pbar Collider Physics\\ Abano Terme, Italy, May 1996}

\end{center}

\end{titlepage}

\section{Introduction}
\subsection{The experiments}\label{sec:exp}

By double hard diffraction it is meant the production of two or 
more jets with a rapidity gap between two jets,
where a rapidity gap is a region in (pseudo)-rapidity 
where no hadrons are produced above a threshold $\mu$.
Double hard diffraction has been 
analysed by the CDF and D0 Collaborations \cite{d0,cdf,d02} at the 
Tevatron $p\,\bar{p}$ collider, operating at $\sqrt{s} = 1.8$ TeV, 
and in photoproduction events by the Zeus Collaboration at the $e\, p$
HERA collider~\cite{zeus}, operating at $\sqrt{s} \simeq 300$ GeV. 
The main difference with respect to
single hard diffraction is the momentum transfer $t$; while in the latter
$t$ is usually much less than 1 $\rm{GeV}^2$, in double hard diffraction
it is very large ($|t|\gsim 10^3\,\rm{GeV}^2$ in the Tevatron 
experiments~\cite{d0,cdf,d02} and $|t|\gsim 30\,\rm{GeV}^2$ at 
HERA~\cite{zeus}). 

In the experiments the events are ranked according to the rapidity 
interval $\Delta\eta$ between the
two leading $E_{\perp}$ jets, and measure the fraction of these events 
which have a gap of width $\Delta\eta_c=\Delta\eta-2R$, with $R$ the
jet-cone size, between the leading $E_{\perp}$ jets. 
One observes~\cite{d0,d02,zeus} a falloff of the gap
fraction as the gap width $\Delta\eta_c$ increases, as expected if the
gaps are merely due to multiplicity fluctuations. However, when $\Delta\eta_c
\gsim 2$ the gap fraction becomes basically independent of $\Delta\eta_c$.
Estimating the contribution of the multiplicity fluctuations by fitting the
high-multiplicity data through a (double) negative binomial distribution (NBD),
the D0 Collaboration~\cite{d02} puts the value of the gap fraction in
excess of the background at
\be
f_g = 1.07 \pm 0.10~(\rm stat) \begin{array}{c} +0.25\\ -0.13 \end{array}
(\rm syst) \%\, .\label{fgap}
\ee
The gap fraction in excess at HERA has been estimated to be about 
7\%~\cite{zeus}.

\subsection{The motivation}\label{sec:mot}

\begin{figure}[htb]
\vspace*{-6.5cm}
\hspace*{-1cm}
\epsfxsize=15cm \epsfbox{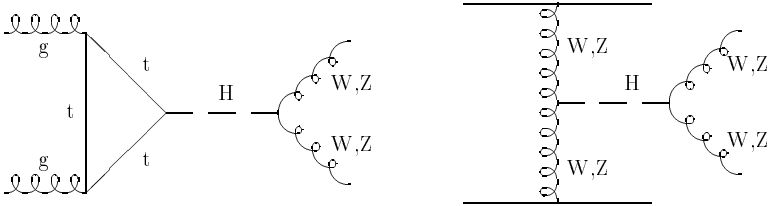}
\vspace*{-11cm}
\caption{Higgs-boson production via ($a$) gluon fusion and ($b$) weak-boson
fusion.}
\label{fig:one}
\end{figure}

The original, and still the main, motivation for analysing double hard 
diffraction events
is to look for a clean signal for the production of a heavy Higgs boson
at hadron supercolliders~\cite{dkt,bj}. A heavy Higgs boson is mainly
produced via gluon fusion, $g\,g\ra H$, mediated by a top-quark loop, 
Fig.~\ref{fig:one}(a). The Higgs boson then decays mainly into a pair of $W$
or $Z$ bosons. Such a signal, though, is going to be swamped by the $W\,W$
QCD and the $t\,\bar t$ backgrounds, Fig.~\ref{fig:two}. Higgs-boson 
production via weak-boson fusion, $W\,W,\; Z\,Z\ra H$, Fig.~\ref{fig:one}(b),
has a smaller rate~\cite{gun} but would have the 
distinctive feature of a rapidity gap in parton production 
since no color is exchanged between the quarks emitting the weak bosons.
Producing a rapidity gap at the parton level is not enough though,
since the gap is usually filled by soft hadrons produced in the rescattering
between the spectators partons in the underlying event.
In addition, the signal might be faked by the exchange of gluons in a
color-singlet configuration~\cite{bj}. Hence the idea to test
the physics of rapidity gaps between jets at the Tevatron.

\section{Double hard diffraction}
\subsection{Gaps at the parton level}\label{sec:doub}

A rapidity gap in parton-parton scattering at the Tevatron
may be produced via $\gamma,\,W,\,Z$-boson exchange in the crossed channel,
$\hat t$. However, this process is concealed by gluon exchange, which is an
${\cal O}(\al_s^2)$ process and whose rate
is bigger by 2-3 orders of magnitude~\cite{bj,che}. Gluon exchange in the
$\hat t$ channel is likely not to produce a gap because the exchanged 
gluon being a color octet radiates off more gluons. Indeed while
in parton-parton scattering with electroweak exchange additional gluon 
bremsstrahlung is mainly in the forward directions, in 
gluon exchange it occurs mainly in the central-rapidity 
region, thus filling the gap~\cite{basic}. 

A gap may also be produced by exchanging two gluons in the $\hat t$ 
channel in a color-singlet configuration. This is an ${\cal O}(\al_s^4)$ 
process, whose rate Bjorken~\cite{bj} estimated to be about 10\% of the 
one-gluon exchange rate, $\hat\sigma_{sing}/\hat\sigma_{oct} \sim 0.1\,$.
The integral of the gluon loop formed by the two-gluon exchange is
dominated by small transverse momenta. Thus the exchange is
given by a hard and a soft gluon. It is not clear though if in this
way a color singlet may be formed because the hard gluon has time
to emit more gluons, the soft gluon that closes the loop being emitted 
much later (or much earlier)~\cite{yuri}.

In the limit of high squared parton c.m. energy $\hat{s}$ and fixed 
$\hat{t}$, an $O(\al_s^5)$ analysis of the gluon-bremsstrahlung
pattern for parton-parton scattering with two-gluon exchange in 
a color-singlet configuration
shows that if the transverse momentum $p_{\perp rad}$ of the 
radiated gluon is of the same order as the transverse momenta $p_{\perp jet}$
of the tagging jets then the gluon is radiated mainly in the 
central-rapidity region, like in the one-gluon exchange case; if
$p_{\perp rad} \ll p_{\perp jet}$, then the gluon is radiated mainly in the
forward direction, i.e. the radiation pattern is similar to
the one of electroweak exchange~\cite{chez}. In other words, if the 
bremsstrahlung
gluon is hard it has a short wavelength and may resolve the color structure
of the two-gluon exchange; if it is soft its resolving power is low and
sees the two exchanged gluons as a color singlet.

In the limit of $\hat{s}\gg |\hat{t}|$ it is possible to resum the leading 
logarithmic contributions, in $\ln(\hat{s}/|\hat{t}|)$, to a scattering 
amplitude to all orders in $\al_s$ by using the Balitsky-Fadin-Kuraev-Lipatov 
(BFKL) theory \cite{fkl}. Resumming the leading virtual radiative corrections
to one-gluon exchange in gluon-gluon scattering, we obtain~\cite{fklo,mt,ddt},
\be
{d\hat\sigma_{oct}\over d\hat{t}} \simeq 
{9\pi\al_s^2\over 2\hat{t}^2}\, \exp\left(-{3\al_s\over\pi}\, \ln{\hat{s}
\over |\hat{t}|}\, \ln{p_{\perp}^2\over\mu^2}\right)\, ,\label{diot} 
\ee
with $p_{\perp}$ the transverse momentum of the outgoing gluons, 
$\hat{t} \simeq - p_{\perp}^2$, and $\mu$ a cutoff that regulates the 
infrared divergence, which is there because we have not included the real 
emissions. Eq.~(\ref{diot}) defines the
scattering as elastic if no soft gluons with $p_{\perp} \gsim \mu$ appear in
the final state. The exponential of eq.~(\ref{diot}) vanishes as $\mu\ra 0$,
which is a general feature of bremsstrahlung emission known as infrared 
catastrophe or Block-Nordsieck mechanism~\cite{ster},
namely it is not possible to produce a gap with one-gluon exchange if 
the resolution of our apparatus is infinitely good. In real life, though, 
there is always a finite contribution to the gap from one-gluon exchange
because we cannot detect soft gluons with $p_{\perp} \lsim \mu$. This 
constitutes a background to the signal we are interested in and at the 
hadron level may be removed by using a double NBD that fits well
the high-multiplicity data~\cite{d02}. In addition, the exponential in
eq.~(\ref{diot}) becomes smaller as the rapidity interval
between the partons $\Delta\eta\simeq\ln(\hat{s}/|\hat{t}|)$ grows,
in qualitative agreement with the data~\cite{d0,d02,zeus}. 

The resummation of the leading virtual radiative corrections
to two-gluon exchange is more problematic, because in this case
the solution of the BFKL equation is well behaved only for the scattering 
between colorless objects~\cite{lip}. For the scattering between partons,
e.g. gluons, the solution~\cite{mt,ddt}
\be
{d\hat\sigma_{sing}\over d\hat{t}} \simeq 
{81\pi^3\al_s^4\over 4\hat{t}^2}\, {\exp\left[24\ln{2}\al_s\, 
\ln(\hat{s}/|\hat{t}|)/\pi\right]\over \left[21\zeta(3)\al_s
\ln(\hat{s}/|\hat{t}|)/2\right]^3}\, ,\label{dinov}
\ee
with $\zeta(3) = 1.20206...$,
is valid only in an asymptotic sense, i.e. as $\Delta\eta\ra\infty$,
since at large but finite $\Delta\eta$ it would exhibit infrared divergences
at each order in the expansion in $\al_s$. Leaving the theoretical
details apart, eq.~(\ref{dinov}) states that even though of higher order 
the signal~(\ref{dinov}) quickly becomes more important 
than the background~(\ref{diot}) as the gap width grows.

\subsection{Gaps at the hadron level}\label{sec:hadr}

In order to make a prediction at the hadron level, eq.~(\ref{diot}) and
(\ref{dinov}) must be convoluted with parton densities, $f(x_{a,b};\mu_s)$,
with $x_{a,b}$ the momentum fractions of the incoming partons and
$\mu_s\gg\lambda_{QCD}$ a factorization scale. 
Thus we may compute the dijet production rate as a function of the rapidity 
difference, $\Delta\eta=\eta_{j_1}-\eta_{j_2}$. In doing that, it is 
convenient to fix the $x$'s, in order to minimize the variations induced 
by the parton densities, which have nothing to do with the parton dynamics
we want to examine~\cite{mt,ddt}. Since $\hat s = x_a x_b s$ and
$\Delta\eta\simeq \ln(\hat s/p_{\perp}^2)$, this implies to run $\Delta\eta$ up
with the collider c.m. energy $\sqrt{s}$, e.g. to compare dijet production
data with a rapidity gap, at fixed values of the $x$'s and the jet
transverse momenta $p_{\perp}$, in the
Tevatron runs at $\sqrt{s} = 1.8$~TeV and $\sqrt{s} = 630$~GeV.
Within the values of $\Delta\eta$ kinematically accessible at the Tevatron, 
the signal~(\ref{dinov}) is not very
sensitive to variations of $\Delta\eta$~\cite{ddt}.

Otherwise in a data sample at fixed $s$ one must run $\Delta\eta$ up
with the $x$'s~\cite{ddt2}. In this case the prediction 
for the gap fraction as a function of the gap width shows an abrupt rise 
at the largest gap widths kinematically allowed.
However much of it is not due to the growth of the singlet contribution in 
the parton dynamics, eq.(\ref{dinov}), but merely to the parton luminosity
which as $x\ra 1$ falls off faster for the inclusive dijet production
than for the one with the gap~\footnote{This kinematic phenomenon is 
exactly the reverse of the one noted for the $K$-factor in inclusive 
dijet production in ref.~\cite{dds}.}.

In the parton densities we choose the factorization scale
$\mu_s\gg\lambda_{QCD}$ because we must allow for the emission of 
soft hadrons in the rescattering between spectator partons in 
the underlying event in accordance with the factorization 
theorems~\cite{css}. In the double diffraction events, it is natural
to identify the scale $\mu_s$ with the threshold $\mu$ above which we
see no hadrons in the rapidity gap. However in the 
experiments~\cite{d0,cdf,d02,zeus} $\mu\simeq\lambda_{QCD}$, thus the 
factorization picture~\cite{css} does not apply, and we need a 
non-perturbative model that lets the gap formed at the parton level survive
the rescattering between the spectator partons, which
would fill the gap with soft hadrons. Using an eikonal model,
Bjorken~\cite{bj} estimated the rapidity-gap survival probability, $<|S^2|>$,
to be about 5-10\%. Combined with the estimate 
$\hat\sigma_{sing}/\hat\sigma_{oct} \sim 0.1$ for the gap production rate
at the parton level~\cite{bj}, this yields a fraction of gaps between jets
at the level of 0.5-1\%, which is in qualitative agreement with 
eq.~(\ref{fgap}).

The gap survival probability, $<|S^2|>$, deals with the low-$p_{\perp}$
physics of the scattering between the two hadrons. It is expected then
to be fairly insensitive to the gap width,
since the rapidity interval between the jets $\Delta\eta$
is a kinematic parameter of the hard-interaction process.
$<|S^2|>$ is expected to decrease as the $s$ increases~\cite{bj,glm},
because the total cross section, $\sigma_{tot}$, is related to the area of the 
soft interactions, $\pi R^2$, and to the unitarity bound by the relation,
$\sigma_{tot} \simeq \pi R^2 \propto \ln{s^2}$. Thus as $s$ increases it is
less and less likely for the two hadrons not to interact.
In addition, $<|S^2|>$ is expected to grow as the momentum fraction $x$ of the 
incoming partons goes to 1, because there is less and less energy available
for the underlying event, i.e. for the spectator partons, in analogy with
the suppression of the underlying event observed in photoproduction 
as $x\ra 1$. Thus if data for dijet production with a gap, at fixed values 
of the $x$'s and the jet transverse momenta $p_{\perp}$, are compared for
Tevatron runs at $\sqrt{s} = 1.8$~TeV and $\sqrt{s} = 630$~GeV,
the main change should come from $<|S^2|>$
and should show an increase in the gap fraction going from 
$\sqrt{s} = 1.8$~TeV to $\sqrt{s} = 630$~GeV.

Finally, we recall that the gap fraction, $f_g\sim 7\%$, in excess of the 
background at HERA~\cite{zeus} is much higher than the one
measured at the Tevatron~(\ref{fgap}). As in $p\,\bar p$ collisions,
in photoproduction in $e\, p$ collisions,
a rapidity gap in dijet production may be formed via electroweak as well as
two-gluon exchange. Having ruled out electroweak exchange because it cannot
account for the size of the gap fraction, two-gluon exchange occurs at
${\cal O}(\al_s^4)$ and receives contributions from resolved and direct
photons. At the parton level singlet exchange in resolved-photon production is 
the same as in sect.~\ref{sec:doub}, and so is the
probability of producing a gap, for equal values of the gap width 
$\Delta\eta_c$~\footnote{Notice, however, that the Tevatron 
experiments~\cite{d0,cdf,d02} use a cone size $R=0.7$, while the ZEUS 
Collaboration~\cite{zeus} uses $R=1$.}; at the hadron level the
gap survival probability, $<|S^2|>$, is expected to be larger because
the $\gamma\, p$ c.m. energy is only a fraction of 
the $e\, p$ c.m. energy, which is $\sqrt{s}\simeq 300$~GeV, and because the
parton densities in the photon are stiffer than the ones in the
proton as $x\ra 1$, and yield a larger contribution for processes where
the underlying event is suppressed. The larger stiffness of the
parton densities in the photon is related to the contribution from 
direct-photon production that beyond the lowest order in dijet production, 
${\cal O}(\al_s)$, cannot be unambiguously disentangled from the
resolved-photon component. The direct-photon component, having no
underlying event, has $<|S^2|>=1$. 

The discussion of the topics in this section, which reflects though the 
state of the art, is very hand-waving and no detailed phenomenological 
predictions have been attempted. The situation may be considerably improved:
at the parton level by performing next-to-leading order (NLO) calculations,
as we now briefly discuss; at the hadron level by dispensing with the 
gap survival probability as we will show in the next section.

\subsection{Improving the model}\label{sec:fut}

The BFKL theory, through which we have derived eq.~(\ref{diot}) and 
(\ref{dinov}), is not suitable for a detailed study of jet
production because within the BFKL theory
the jets have no structure, i.e. are point-like.
This drawback is even more acute for dijet production with a rapidity gap
because as we said in the Introduction the experiments measure the gap width, 
$\Delta\eta_c$, between the edges of the jet cones,
which differs from the rapidity difference, $\Delta\eta$, between the 
jet centers by the cone sizes $R$, $\Delta\eta_c = \Delta\eta - 2R$. 
The BFKL approximation is not able to distinguish between $\Delta\eta$ 
and $\Delta\eta_c$. In addition, as we pointed out the calculation for 
eq.~(\ref{dinov}) is not infrared stable.

In order to examine the gap fraction as a function of the gap width between
the jet-cone edges, while accounting properly for the cone structures, we
need an exact higher-order calculation which includes, though, the basic 
features of color-singlet exchange. The simplest calculation of this kind 
for the dijet production rate with a rapidity gap is ${\cal O}(\al_s^4)$.
At the moment this is unfeasible because one needs to know
two-loop matrix elements, which have not been computed yet.
However the gap fraction, i.e. the ratio of the rapidity-gap to
the inclusive dijet production, may be computed subtracting out from the
unity the ratio of the three-jet to the inclusive dijet 
production, for which at ${\cal O}(\al_s^4)$ we need only known one-loop
and tree-level matrix elements. The preparation of an NLO three-jet
Monte Carlo, which could also serve to this scope, is in 
progress~\cite{kunszt}.
The jet production with a gap in rapidity would then be computed by
requiring that any extra partons besides the ones we tag on be emitted
within the jet cones. Therefore a distinction between octet and singlet
contributions would not be done. In addition, the calculation would be
infrared stable, and the dependence on the factorization scale
strongly reduced.

\section{Rapidity-gap physics at the LHC}
\subsection{Higgs-boson production via weak-boson fusion}

\begin{figure}[htb]
\vspace*{-6.5cm}
\hspace*{-1cm}
\epsfxsize=15cm \epsfbox{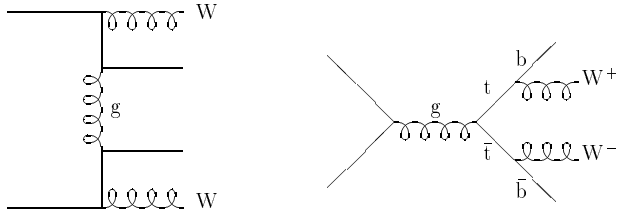}
\vspace*{-11.5cm}
\caption{($a$) One of the Feynman diagrams that contribute to
$W\,W$ QCD production, and ($b$) $t\,\bar t$ production.}
\label{fig:two}
\end{figure}

As we anticipated in the Introduction, double hard diffraction
offers a test ground for the production
of a Higgs boson via weak-boson fusion, $W\,W,\; Z\,Z\ra H$, 
at hadron supercolliders~\cite{dkt,bj}. Differently from the
$g\,g\ra H$ production mechanism, Fig.~\ref{fig:one}(a), 
$W\,W,\; Z\,Z\ra H$ production, Fig.~\ref{fig:one}(b), has a 
distinct radiation pattern with a gap in parton production in the
central-rapidity region, because no color is exchanged 
between the quarks that emit the weak bosons.
This may allow us to distinguish the $W\,W,\; Z\,Z\ra H$ signal from the
overwhelming $W\,W$ QCD and $t\,\bar t$ backgrounds,
Fig.~\ref{fig:two}, which would have no gaps in parton production.

As we have seen in sect.~\ref{sec:hadr} it is not sufficient to produce 
a gap at the parton level, since the soft hadrons produced by
the underlying event of the
$p\, p$ scattering usually fill the gap. This scenario applied as such
at the SSC collider, and led Bjorken~\cite{bj} to introduce the gap 
survival probability, $<|S^2|>$. However, at the LHC collider the
requirement of running at very high luminosity yields
overlapping events in the same bunch crossing which are an additional
source of soft hadrons and would further, and hopelessly, suppress the 
gap signal. A way out of this deadlock is to require a gap in minijet 
production~\cite{dks,bpz} rather than in soft-hadron production~\footnote{A
good introduction to the topic is in ref.~\cite{zepp}.}.
This has the additional advantage of dispensing with the
gap survival probability, because the production of soft hadrons
in the rescattering of the spectator partons is clearly unrestricted.
Then $<|S^2|> = 1$, and since $\mu_s\gg\lambda_{QCD}$ there is no
factorization breaking~\cite{ddt}.

\subsection{Forward-jet tagging and the minijet veto}

A heavy Higgs boson decays predominantly into a pair of $W$ or $Z$ bosons,
Fig.~\ref{fig:one}.
In what follows we assume the subsequent leptonic decay of the weak 
bosons~\footnote{In the case of hadronic decay,
the rapidity-gap window is partially colored by the decay products of
the weak bosons, however if the Higgs boson is very heavy the outgoing
weak bosons are far off shell and the hadronic decay products are boosted
into cones of small opening angle, which should not spoil the
rapidity-gap signature~\cite{bj}.}, and follow the outline of ref.~\cite{bpz}.
The incoming $W$'s or $Z$'s tend to
have a small momentum fraction, $x_{W,Z}\ll 1$, of the parent quarks, however 
they must be energetic enough to produce the Higgs boson, $E_{W,Z}\gsim m_H/2$.
This implies that the outgoing quarks, that
carry the momentum fraction $(1-x)$ and usually hadronize into jets,
are very energetic. In addition, the incoming-$W,Z$ propagators yield
the largest contribution when the transverse momentum, $p_{\perp j}$, 
of the incoming $W$'s or $Z$'s and of the outgoing jets is $p_{\perp j}\lsim
m_{W,Z}$. Therefore the outgoing jets tend to be energetic and to come out
at a small angle, i.e. in the forward-rapidity region. Requiring then a
forward-jet tagging greatly reduces the $W\,W$ QCD and $t\,\bar t$ 
backgrounds with little loss for the signal. 
 
Since the Higgs boson is heavy, the transverse momenta of the outgoing $W$'s 
or $Z$'s, and hence of the leptons the $W$'s or $Z$'s decay into, are large.
Thus the leptons usually come out in the central-rapidity region. An 
additional background reduction is achieved by requiring that the leptons
are central and that they are separated by a large rapidity gap from
the forward jets.

However, a further large background suppression, and the one which
makes eventually the signal to stand out, is the very essence of the
rapidity-gap physics: the signal and the background have a very different
radiation pattern; namely for the background color is exchanged in the
crossed channel, and so the transverse scale of the color exchange is
is of the order of the hardness of the process, tipically $Q= {\cal O}$(1 TeV).
If we assume that the probability of parton bremsstrahlung is
given by $f_s=\al_s \ln(Q^2/p_{\perp min}^2)$ then multiple bremsstrahlung
occurs when $f_s \simeq 1$, i.e. when $p_{\perp min} \sim$~50 GeV.
Thus we expect the background to have multiple minijet emission in the 
50~GeV range~\footnote{The minijet activity may be preliminarly tested at
the Tevatron, by considering the minijet emission at ${\cal O}(\al_s^3)$
in high-$p_{\perp}$ dijet events~\cite{sz}.}.
On the other hand in the signal there is no color emission
but in the formation of the outgoing quarks. As we have seen in the paragraph
above the transverse scale associated with that is $p_{\perp j}= Q \sim$~50-80
GeV, thus for the signal multiple minijet emission is expected to occur when
$p_{\perp min} \sim$~5 GeV. Vetoing the production of
minijets with, say, $p_{\perp min} \gsim$~20 GeV  drastically reduces the
background while affecting the signal very little. 

At the LHC a gauge of the efficiency of the background-reduction techniques
described above is $Z+2$-jet production. At the parton level the signal
is given by quark scattering, $q\, q\ra q\, q\, Z$, via $\gamma\,,W\,,Z$-boson
exchange, and resembles the Higgs-boson production via weak-boson fusion
described above. The background is given by $Z+2$-jet events in 
${\cal O}(\al_s^2)$ Drell-Yan production. Considering the leptonic decay
of the $Z$ boson and implementing the forward-jet and central-lepton tagging 
and the minijet veto cuts as above reduces the huge background to a level
below the signal~\cite{rsz}.

\section{Conclusions}

We have reviewed the state of the art for the production of rapidity gaps
between jets at the Tevatron and HERA colliders, which may be accounted for 
by assuming the exchange of a color singlet in the crossed channel, with
suppressed radiation in the central-rapidity region. We have described
how at LHC energies this translates into a suppression of minijet emission
which may be used to enhance Higgs-boson production via electroweak-boson
exchange over the dominant QCD backgrounds.

\section*{Acknowledgments}
I wish to thank bj Bjorken, Andrew Brandt, Jon Butterworth, Yuri Dokshitzer,
Lonya Frankfurt,
Bob Hirosky, David Kosower, Brent May and Phil Melese for contributing 
to my understanding of this topics, and the organization of the Workshop
for the hospitality and the support.


\end{document}